\documentclass[aps,pra,twocolumn,nofootinbib, superscriptaddress,10pt]{revtex4-1}

\usepackage{graphicx}
\usepackage{epstopdf}
\usepackage{amsmath}
\usepackage{amssymb}
\usepackage{mathrsfs}
\usepackage{amsthm}
\usepackage{bm}
\usepackage{url}
\usepackage[T1]{fontenc}
\usepackage{csquotes}
\MakeOuterQuote{"}

\newtheoremstyle{note}
  {\topsep/2}               
  {\topsep/2}               
  {}                      
  {\parindent}            
  {\itshape}              
  {.}                     
  {5pt plus 1pt minus 1pt}
  {}

\theoremstyle{note}
\newtheorem{theorem}{Theorem}
\newtheorem{lemma}{Lemma}

\newtheorem{corollary}{Corollary}
\newtheorem{proposition}{Proposition}

\theoremstyle{definition}

\theoremstyle{remark}
\newtheorem{remark}{Remark}

\newcommand{\tr}{\operatorname{tr}}

 \newcommand{\rmi}{\mathrm{i}}

 \newcommand{\rmA}{\mathrm{A}}
 \newcommand{\rmB}{\mathrm{B}}

 \newcommand{\rmF}{\mathrm{F}}

 \newcommand{\rmR}{\mathrm{R}}

 \newcommand{\caB}{\mathcal{B}}
 
 \newcommand{\caH}{\mathcal{H}}

 \newcommand{\bfq}{\mathbf{q}}

  \newcommand{\scrB}{\mathscr{B}}
 
 \newcommand{\scrO}{\mathscr{O}}
 \newcommand{\scrP}{\mathscr{P}}
 \newcommand{\scrS}{\mathscr{S}}

\newcommand{\be}{\begin{equation}}
\newcommand{\ee}{\end{equation}}
\newcommand{\ba}{\begin{align}}
\newcommand{\ea}{\end{align}}

\def\<{\langle}  
\def\>{\rangle}  

\def\eqref#1{\textup{(\ref{#1})}}  
\newcommand{\eref}[1]{Eq.~\textup{(\ref{#1})}}

\newcommand{\thref}[1]{Theorem~\ref{#1}}
\newcommand{\Thref}[1]{Theorem~\ref{#1}}
\newcommand{\thsref}[1]{Theorems~\ref{#1}}

\newcommand{\lref}[1]{Lemma~\ref{#1}}
\newcommand{\Lref}[1]{Lemma~\ref{#1}}
\newcommand{\lsref}[1]{Lemmas~\ref{#1}}

\newcommand{\pref}[1]{Proposition~\ref{#1}}
\newcommand{\Pref}[1]{Proposition~\ref{#1}}
\newcommand{\psref}[1]{Propositions~\ref{#1}}

\newcommand{\crref}[1]{Corollary~\ref{#1}}
\newcommand{\Crref}[1]{Corollary~\ref{#1}}
\newcommand{\crsref}[1]{Corollaries~\ref{#1}}

\newcommand{\cref}[1]{Conjecture~\ref{#1}}
\newcommand{\Cref}[1]{Conjecture~\ref{#1}}

\newcommand{\rcite}[1]{Ref.~\cite{#1}}

\begin{document}
\title{Zero Uncertainty States in the Presence of  Quantum Memory}

\author{Huangjun Zhu}
\email{zhuhuangjun@fudan.edu.cn}

\affiliation{State Key Laboratory of Surface Physics and Department of Physics, Fudan University, Shanghai 200433, China}

\affiliation{Institute for Nanoelectronic Devices and Quantum Computing, Fudan University, Shanghai 200433, China}

\affiliation{Center for Field Theory and Particle Physics, Fudan University, Shanghai 200433, China}

\affiliation{Collaborative Innovation Center of Advanced Microstructures, Nanjing 210093, China}

\begin{abstract}
The uncertainty principle imposes a fundamental limit on  predicting the measurement outcomes of incompatible observables even if  complete classical information of the system state is known. The situation is  different if one can build a quantum memory entangled with 
 the system. Zero uncertainty states (in contrast with minimum uncertainty states) are peculiar quantum states that can eliminate uncertainties of incompatible von Neumann observables once assisted by suitable measurements on the memory.
 Here we determine all  zero uncertainty states of any given set of nondegenerate  observables and determine the minimum entanglement required. It turns out all zero uncertainty states are maximally entangled in a generic case, and vice versa,  even if these observables are only weakly incompatible. Our work establishes a simple and precise connection between zero uncertainty and maximum entanglement, which is of interest to  foundational studies and practical applications, including quantum certification and verification. 
\end{abstract}

\date{\today}
\maketitle

\section{INTRODUCTION}
The uncertainty principle represents a key distinction between quantum mechanics and classical mechanics and is still a focus of current research \cite{Heis27,BuscLW14,ColeBTW17,WernF19}.
It imposes a fundamental limit on our ability to predict the measurement outcomes of incompatible observables, such as position and momentum \cite{Kenn27,Robe29}. 
However,  uncertainty relations  have to be modified in the presence of a quantum memory because entanglement between the memory and  system can  reduce
the uncertainty \cite{EinsPR35,BertCCR10,LiXXL11,PrevHCF11,TomaR11,ColeCYZ12}. Besides foundational significance, this simple fact is of  interest to diverse applications, including entanglement detection \cite{BertCCR10,LiXXL11,PrevHCF11} and quantum cryptography \cite{BertCCR10,TomaR11,TomaLGR12}. Nevertheless, several fundamental questions are still left open. Notably, what quantum states of the system and memory  can minimize or even eliminate the uncertainty completely?  How much entanglement is required to achieve this goal?

In this paper  we are interested in those quantum states that  can  eliminate the uncertainty completely, which are referred to as
\emph{zero uncertainty states} (ZUSs) henceforth. In contrast with  the familiar minimum uncertainty states and intelligent states  studied in the literature \cite{Stol70,HillM93,AragCS76}, our definition of ZUSs does not rely on a specific uncertainty measure, but has a clear operational interpretation. 
In particular, we determine all ZUSs with respect to any given set of nondegenerate observables on a finite-dimensional Hilbert space coupled with a quantum memory. We also  determine the minimum entanglement required to construct any ZUS. It turns out all ZUSs are determined by a simple graph constructed from transition probabilities between eigenbases of the relevant observables. Moreover, all ZUSs are \emph{maximally entangled states} (MESs), and vice versa, whenever this graph is connected. 
Notably, this is the case for a generic set of two or more  observables, even if these observables are only weakly incompatible. Nevertheless, ZUSs and MESs are not necessarily pure states.

Our study establishes a simple and  precise connection between  zero uncertainty and maximum entanglement, which is independent of specific uncertainty and entanglement measures. Moreover, our approach applies to arbitrary sets of nondegenerate observables,  in sharp contrast with most previous approaches, which are restricted to two observables or complementary observables. 
This work 
may shed light on the  foundational studies of uncertainty relations, quantum entanglement, and quantum steering \cite{EinsPR35,WiseJD07,SupiH16,CavaS17,UolaCNG20}. Meanwhile, it is of direct interest to
many tasks in   quantum information processing, including remote state preparation \cite{BennDSS01,BerrS03} and semi-device-independent quantum certification and verification \cite{MayeY04,SupiH16,MccuPBM16,SupiB20,EiseHWR20}.

\section{RESULTS}
\noindent\textbf{Maximally entangled states}\\
To establish our main results, first we need to better understand MESs. A bipartite state $\rho$ on the Hilbert space  $\caH_\rmA\otimes \caH_\rmB$ of dimension $d_\rmA\times d_\rmB$ is a MES if we can create 
 every  state on $\caH_\rmA\otimes \caH_\rmB$ from $\rho$ by local operations and classical communication (LOCC)~\cite{HoroHHH09}. This definition is independent of any specific entanglement measure and is thus quite appealing to the current study.
 When $d_\rmA\leq d_\rmB$, the state 
$|\Phi\>:=\sum_{j=0}^{d_\rmA-1}|jj\>/\sqrt{d_\rmA}$ is 
a canonical example, where $\{|j\>\}_{j=0}^{d_\rmA-1}$ and $\{|j\>\}_{j=0}^{d_\rmB-1}$ are the computational bases of $\caH_\rmA$ and $\caH_\rmB$, respectively.   A MES is not necessarily  pure, as clarified in the following lemma, essentially proved in  \rcite{LiZFF12}; see the Supplementary Information  for an independent proof. 
\begin{lemma}\label{lem:MES}
	Let $\rho$ be a bipartite state on $\caH_\rmA\otimes \caH_\rmB$ with $d_\rmA\leq d_\rmB$. Then the following statements are equivalent. 
	\begin{enumerate}
		\item $\rho$ is a MES.
		\item $H(\rmA|\rmB)_\rho=-\log_2 d_\rmA$.
		\item $E_\rmR(\rho)=\log_2 d_\rmA$. 
		\item $E_\rmF(\rho)=\log_2 d_\rmA$. 
		\item $\rho$ has a spectral decomposition $\rho=\sum_s \lambda_s |\Psi_s\>\<\Psi_s|$ such that all $|\Psi_s\>$ are MESs,  and
		$\tr_\rmA(|\Psi_s\>\<\Psi_s|)$ have mutually orthogonal supports.
	\end{enumerate}
\end{lemma}
Here $H(\rmA|\rmB)_\rho=S(\rho)-S(\rho_\rmB)$ is the conditional entropy of A given B, where $S(\rho)$ and $S(\rho_\rmB)$ are the von Neumann entropies of $\rho$ and  $\rho_\rmB:=\tr_\rmA(\rho)$,  respectively. $E_\rmR(\rho)$ and $E_\rmF(\rho)$ are the relative entropy of entanglement and   entanglement of formation \cite{HoroHHH09}. 
By \lref{lem:MES}, all MESs on $\caH_\rmA\otimes \caH_\rmB$ are pure when $d_\rmA\leq d_\rmB<2d_\rmA$ \cite{LiZFF12}. In general, each MES is a convex mixture of pure MESs whose local supports for Bob are mutually orthogonal. Given a MES $\rho$ on $\caH_\rmA\otimes \caH_\rmB$ with $d_\rmA\leq d_\rmB$, let $\caH_{\rmB'}$ be the support of $\rho_\rmB$.  
Then $\caH_{\rmB'}$ has a decomposition $\caH_{\rmB'}=\caH_{\rmB_1}\otimes \caH_{\rmB_2}$ with $\dim(\caH_{\rmB_1})=d_\rmA$ 
such that $\rho=|\Phi'\>\<\Phi'|\otimes \tau$, where $|\Phi'\>$ is a pure MES in $\caH_\rmA\otimes \caH_{\rmB_1}$, and $\tau$ is a full-rank density operator on $\caH_{\rmB_2}$. So  all MESs on $\caH_\rmA\otimes \caH_\rmB$ are equivalent under local operations of Bob.
\begin{corollary}\label{cor:MES}
	All MESs on $\caH_\rmA\otimes \caH_\rmB$ with $d_\rmA\leq d_\rmB$ can be turned into each other by local operations on $\caH_\rmB$. When $d_\rmA\leq d_\rmB<2d_\rmA$, all MESs are pure and can be turned into each other by  unitary transformations on $\caH_\rmB$. 
\end{corollary}

\bigskip
\noindent\textbf{Zero uncertainty states}\\
Consider the uncertainty game in which Alice can  measure  $m$ nondegenerate von Neumann observables   $\scrO=\{O_x\}_{x=1}^m$ on $\caH_\rmA$ with uniform probabilities (generalization to nonuniform probabilities is straightforward),  and Bob is asked to predict the measurement outcome given the specific observable chosen by Alice \cite{BertCCR10}. Let $\caB_x=\{|\psi_{x k}\>\}_k$ be an orthonormal eigenbasis of $O_x$  and $\scrB=\{\caB_x\}_{x=1}^m$. Then  predicting the outcome of $O_x$ amounts to predicting the  outcome of the projective measurement on the basis $\caB_x$.  When these observables are incompatible (do not commute with each other), in general Bob cannot predict the measurement outcome with certainty even if he knows the complete classical description of the system state as characterized by the density matrix $\rho_\rmA$. In the case of two observables for example, the uncertainties of the measurement outcomes satisfy the 
Maassen-Uffink inequality \cite{MaasU88},
\begin{equation}\label{eq:UCRMU}
H(O_1)+H(O_2)\geq \log_2(c^{-1}),
\end{equation}
where $H(O_1)$ and $H(O_2)$ are the Shannon entropies of the measurement outcomes of $O_1$ and $O_2$, respectively, and $c=\max_{j,k}|\<\psi_{1j}|\psi_{2k}\>|^2$.
\begin{figure}
	\includegraphics[width=7.5cm]{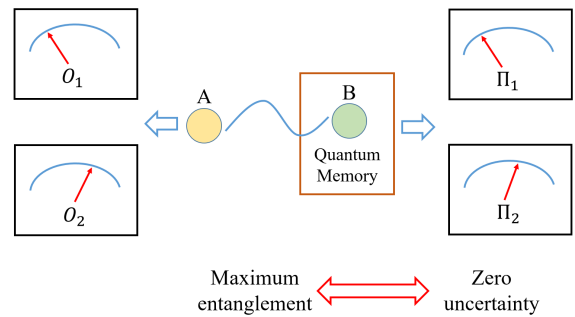}
	\caption{Connection between zero uncertainty and maximum entanglement in the uncertainty game. 
		By performing suitable measurements (depending on the observables of Alice) on his quantum memory, Bob can better predict the measurement outcomes of Alice. 	When the set of observables of Alice is irreducible, Bob  can predict these measurement outcomes  with certainty iff his quantum memory is maximally entangled with Alice.	
	}
\end{figure}

The situation is different if Bob holds a quantum memory with Hilbert space $\caH_\rmB$ and can create an entangled state $\rho$ on the joint system $\caH_\rmA\otimes \caH_\rmB$, as illustrated in Fig.~1.
Suppose Alice  chooses the basis $\caB_x$ (observable $O_x$), then   Bob can perform a generalized measurement characterized by a positive operator-valued measure (POVM) $\{\Pi_{x k}\}_k$ on  his subsystem  $\caH_\rmB$, where $\Pi_{x k}$ corresponds to guessing the outcome $k$ given the measurement basis $\caB_x$ of Alice. The average success guessing probability reads $\sum_x p_x/m$ with
\begin{equation}\label{eq:GuessProb}
p_x=\sum_k \tr[\rho(|\psi_{x k}\>\<\psi_{x k}| \otimes \Pi_{x k} )]=\sum_k\tr(\rho_{x k}\Pi_{x k}),
\end{equation}
where  $\rho_{x k}=\<\psi_{x k}|\rho|\psi_{x k}\>$ are subnormalized reduced states  of Bob. Note that $p_x$ is also the probability that the POVM $\{\Pi_{x k}\}_k$ can successfully distinguish the ensemble of states 
\begin{equation}\label{eq:ensemble}
\scrS(\rho,\caB_x)=\{\<\psi|\rho|\psi\>\,:\,|\psi\>\in \caB_x\}.
\end{equation}
The maximum of the average guessing probability  over all POVMs can be determined by semidefinite programming, and this maximum  is determined by the state $\rho$ and the basis set $\scrB$ (or the observable set $\scrO$) of Alice.

Given a set of observables $\scrO=\{O_x\}_{x=1}^m$ or bases $\scrB=\{\caB_x\}_{x=1}^m$ for Alice, 
a joint state $\rho$ of Alice and Bob is a \emph{zero uncertainty state} (ZUS) if Bob 
can predict the measurement outcome of Alice with certainty  by a suitable measurement depending on the choice of Alice. Given a ZUS,  the  guessing probability $p_x$ for each measurement of Alice can attain the maximum  1, and the conditional entropy $H(O_x|\rmB)$ is 0.  In contrast with   minimum uncertainty states and intelligent states \cite{Stol70,HillM93,AragCS76}, ZUSs not only minimize the uncertainty, but also eliminate the uncertainty completely. Moreover, here the definition is independent of any specific uncertainty measure.

To appreciate the significance of entanglement to constructing a ZUS, consider an example with two observables, in which case the uncertainty relation in \eref{eq:UCRMU} is modified as follows \cite{BertCCR10},
\begin{equation}\label{eq:UCRQM}
H(O_1|\rmB)+H(O_2|\rmB)\geq \log_2(c^{-1})+H(\rmA|\rmB)_\rho.
\end{equation}
Here the  conditional entropy $H(\rmA|\rmB)_\rho$  manifests the impact of entanglement in reducing the uncertainty. Interestingly, a variant of \eref{eq:UCRQM} may  be interpreted as uncertainty-reality complementarity, which builds on an intimate connection between uncertainty (or rather certainty) and reality \cite{EinsPR35,BiloA15,DiegA18,Rudn18}.

By \eref{eq:UCRQM}, any ZUS $\rho$ must satisfy the inequalities
\begin{equation}\label{eq:ZUScEntropy}
E_\rmF(\rho)\geq E_\rmR(\rho)\geq -H(\rmA|\rmB)_\rho \geq \log_2(c^{-1}),
\end{equation}
where the second one is derived in  \rcite{PlenVP00} (cf.~\rcite{ZhuHC17C}).
Suppose  $O_1$ and $O_2$ are complementary, so that  $\caB_1$ and $\caB_2$ are mutually unbiased, which means $|\<\psi_{1j}|\psi_{2k}\>|^2=1/d_\rmA$ for all $j,k$ \cite{DurtEBZ10}. Then we have $c=1/d_\rmA$ and 
\begin{equation}
E_\rmF(\rho)= E_\rmR(\rho)=-H(\rmA|\rmB)_\rho =\log_2 d_\rmA,
\end{equation}
 which implies that $d_\rmB\geq d_\rmA$ and  $\rho$ is a MES by \lref{lem:MES}, in which case $\rho$ is indeed a ZUS.  Unfortunately, this reasoning does not work in general.  To address this problem, we need a completely different line of thinking.

\bigskip

\noindent\textbf{Key observations about  zero uncertainty states}\\
Bob can predict the measurement outcome on the basis $\caB_x$
with certainty iff the ensemble $\scrS(\rho,\caB_x)$ defined in \eref{eq:ensemble} is perfectly distinguishable, that is, all states in  $\scrS(\rho,\caB_x)$ have mutually orthogonal supports. The state $\rho$ is a ZUS with respect to $\scrB=\{\caB_x\}_{x=1}^m$ iff each ensemble $\scrS(\rho,\caB_x)$  is perfectly distinguishable. The following three propositions are simple corollaries of these observations.

\begin{proposition}\label{pro:ZUSsupport}
Suppose $\rho$ is a ZUS, then any state supported in the support of $\rho$ is  a ZUS. 	
\end{proposition}

\begin{proposition}\label{pro:ZUSsum}
	Suppose  $\rho_1$ and $\rho_2$ are two ZUSs on $\caH_\rmA\otimes \caH_\rmB$. If $\tr_\rmA(\rho_1)$ and $\tr_\rmA(\rho_2)$ have orthogonal supports, then any convex mixture of $\rho_1$ and $\rho_2$ is  a ZUS. 
\end{proposition}

\begin{proposition}\label{pro:ZUSCPTP}
	Suppose $\rho$ is bipartite state on $\caH_\rmA\otimes \caH_\rmB$ and 
	$\Lambda$ is a completely positive and trace-preserving (CPTP) map (quantum channel)  from system B to system $\tilde{\rmB}$. Then  $\rho$ is a ZUS  if $(1\otimes \Lambda)(\rho)$ is. 
\end{proposition}
Here \psref{pro:ZUSsupport} and \ref{pro:ZUSsum} are tied to the fact that mixture of quantum states can only reduce distinguishability unless the reduced states of Bob have orthogonal supports. \Pref{pro:ZUSCPTP} follows from the simple fact that quantum operations cannot enhance distinguishability.  Two states $\rho_1$ and $\rho_2$ on $\caH_\rmA\otimes \caH_\rmB$ are \emph{equivalent} if they can be turned into each other by local operations on $\caH_\rmB$. In that case, $\rho_1$ is a ZUS with respect to $\scrB$ iff $\rho_2$ is. Under these local operations,  ZUSs  divide into  equivalent classes. 

\bigskip

\noindent\textbf{Transition graphs}\\
To determine ZUSs with respect to a given basis set $\scrB=\{\caB_x\}_{x=1}^m$ in $\caH_\rmA$, we first  need to pinpoint a key property of the basis set.  The \emph{transition graph} $G(\scrB)$ of  $\scrB$  is an $m$-partite graph with $md_\rmA$ vertices which are in one-to-one correspondence with the basis states (identical states in different bases correspond to different vertices).  Two different vertices are adjacent iff the corresponding states are not orthogonal, that is, the transition probability between the two states is nonzero. The graph $G(\scrB)$ encodes the incompatibility structure of the basis set $\scrB$, which is crucial to studying ZUSs and quantum verification, as we shall see later.

\begin{figure}
	\includegraphics[width=5.5cm]{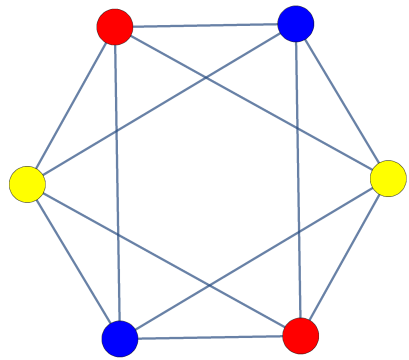}
	\caption{Transition graph of three mutually unbiased bases for a qubit, which correspond to the  eigenbases of the three Pauli matrices. Two vertices of the same color are associated with the two states in the same basis. This transition graph is connected, so the corresponding basis set is irreducible.
	}
\end{figure}

The basis set $\scrB$ is \emph{irreducible} if  the transition graph $G(\scrB)$ is connected, in which case the projectors onto basis states generate the whole operator algebra on $\caH_\rmA$. Any basis set composed of $m\geq2$ mutually unbiased bases (or generic random bases)  is irreducible since the  transition graph is a complete $m$-partite  graph, as illustrated in Fig.~2.

The basis set $\scrB$ is \emph{reducible} if the transition graph $G(\scrB)$ is not connected. In this case, the basis set may be seen as a direct sum of basis sets defined on smaller subspaces. Suppose $G(\scrB)$ has $g$ connected components $G_1, G_2, \ldots, G_g$, then each component $G_a$ is also an $m$-partite graph in which all parties have the same number of  vertices.   Let $\caB_x^a$ be the subset of $\caB_x$ that corresponds to the vertices in the component $G_a$. The \emph{component subspace} $\caH_{\rmA,a}$ associated with  component $a$ is the subspace of $\caH_\rmA$ spanned by  all $|\psi\>\in \caB_x^a$, with 
\emph{component projector} and \emph{component rank} given by
\begin{align}\label{eq:ComponentProj}
P_a=P_a(\scrB)=\sum_{|\psi\>\in \caB_x^a } |\psi\>\<\psi|,\quad 
r_a=\tr(P_a). 
\end{align}
These definitions are independent of the choice of the basis index $x$. Denote by  $\scrP(\scrB):=\{P_a(\scrB)\}_{a=1}^g$ the set of component projectors, which are mutually orthogonal. In this way, $\caH_\rmA$ decomposes into a direct sum of component subspaces $\caH_{\rmA,a}$.  In addition, $\caB_x^a$ for $x=1,2,\ldots, m$ can be regarded as bases in $\caH_{\rmA,a}$, and  the basis set  $\scrB^a: =\{\caB_x^a\}_{x=1}^m$  is irreducible for $\caH_{\rmA,a}$.

\bigskip

\noindent\textbf{Connect zero uncertainty states with maximally entangled states}\\
Now we are ready to  present our main results on  ZUSs and MESs as illustrated in Fig.~1. First, we  clarify when Bob can predict the outcome of one projective measurement of Alice. The following lemma proved in the Supplementary Information  is  a stepping stone to understanding ZUSs in the presence of a quantum memory.
\begin{lemma}\label{lem:PerfectDist}
	Suppose $\rho=|\Psi\>\<\Psi|$ is a bipartite pure state on $\caH_\rmA\otimes \caH_\rmB$ and $\caB$ is an orthonormal basis for $\caH_\rmA$. Then the ensemble $\scrS(\rho, \caB)$ is perfectly distinguishable iff $\caB$ is an eigenbasis of $\rho_\rmA=\tr_\rmB(\rho)$. 
\end{lemma}
Note that Bob can predict the measurement outcome of Alice on the basis $\caB$ with certainty iff the ensemble $\scrS(\rho, \caB)$ defined in \eref{eq:ensemble} is perfectly distinguishable. By \lref{lem:PerfectDist}, this is the case iff $\rho_\rmA$ is diagonal with respect to the basis $\caB$. Therefore, the pure state $\rho$ is a ZUS with respect to a basis set $\scrB$ iff $\rho_\rmA$ is diagonal with respect to each basis in $\scrB$. When  $\scrB$ is irreducible, it turns out only MESs can satisfy this condition. Based on this observation we can derive the following theorem as proved in the Supplementary Information.
\begin{theorem}\label{thm:ZUS}
	Suppose $\scrB$ is an irreducible basis set in $\caH_\rmA$. Then a bipartite (pure or mixed) state $\rho$ on  $\caH_\rmA\otimes \caH_\rmB$  is a ZUS with respect to $\scrB$ iff  $d_\rmB\geq d_\rmA$ and $\rho$ is a MES.  
\end{theorem}
By \thref{thm:ZUS} and \lref{lem:MES}, each ZUS with respect to $\scrB$ is a tensor product of a pure MES and  an ancillary state. In addition, to attain zero uncertainty, the measurements of Bob on the support of $\rho_\rmB$ are  uniquely determined by the counterpart of Alice, as shown in
Supplementary note E. All ZUSs with respect to $\scrB$ can be turned into each other by local operations on $\caH_\rmB$ and thus form a single equivalent class.  If $d_\rmA\leq d_\rmB<2d_\rmA$, then all ZUSs are pure and can be turned into each other by  unitary transformations on $\caH_\rmB$. These results hold as long as the transition graph $G(\scrB)$ is connected, even if $\scrB$ consists of only two nearly identical bases, so that the corresponding observables are only weakly incompatible, as quantified by the commutator or incompatibility robustness \cite{HeinKR15,Haap15,UolaBGP15}.  

\Thref{thm:ZUS} establishes a simple and precise connection between zero uncertainty and maximum entanglement, which is independent of  specific uncertainty and entanglement measures. This connection offers a fresh perspective for understanding  the uncertainty principle in the presence of a quantum memory \cite{BertCCR10}. It may also shed light on uncertainty-reality complementarity given the close relation between the notation of uncertainty and that of reality \cite{EinsPR35,BiloA15,DiegA18,Rudn18}.

\bigskip
\noindent\textbf{Zero uncertainty states with respect to reducible basis sets}\\
Next, we determine  ZUSs with respect to a reducible basis set $\scrB$.
\begin{theorem}\label{thm:ZUS2}
	Suppose $\scrB$ is a set of orthonormal bases in $\caH_\rmA$ and has $g$ irreducible components with
	component subspaces $\caH_{\rmA,a}$, component projectors $P_a$, and component ranks $r_a$ for $a=1,2,\ldots, g$. Let $\rho$ be a bipartite state on  $\caH_\rmA\otimes \caH_\rmB$ and $\rho_a=(P_a\otimes 1_\rmB)\rho(P_a\otimes 1_\rmB)$.   Then $\rho$  is a ZUS with respect to $\scrB$ iff	the following three conditions hold: $r_a\leq d_\rmB$ whenever $\tr(\rho_a)>0$;
	each $\rho_a$ with  $\tr(\rho_a)>0$ is a (subnormalized) MES on $\caH_{\rmA,a}\otimes \caH_\rmB$; all $\tr_\rmA(\rho_a)$  have mutually orthogonal supports.
\end{theorem}
\Thref{thm:ZUS2} follows from \thref{thm:ZUS}. Recall that  the basis set $\scrB$ can be regarded as  a direct sum of irreducible basis sets $\scrB^a$ defined on component  subspaces $\caH_{\rmA,a}$.  So $\rho$  is a ZUS with respect to $\scrB$ iff	its restriction $\rho_a$ on $\caH_{\rmA,a}\otimes \caH_\rmB$   is a ZUS with respect to $\scrB^a$ and,  in addition, all $\tr_\rmA(\rho_a)$  have mutually orthogonal  supports. Note that coherence between different component subspaces are useless to constructing ZUSs.
In addition, 
the dimension  $d_\rmB$ of $\caH_\rmB$ must satisfy $d_\rmB\geq r_{\min}$ in order to construct a ZUS, where $r_{\min}=\min_{1\leq a\leq g} r_a$ is the minimum component rank. When the lower bound is saturated,  every ZUS is a pure MES on $\caH_{\rmA,a}\otimes \caH_\rmB$, where $\caH_{\rmA,a}$ is a component subspace of dimension $r_{\min}$. Furthermore, $r_{\min}$  determines the minimum entanglement required to construct a ZUS, as shown in the Methods section.

In the case of pure states, ZUSs admit a much simpler characterization, as shown in 
the following corollary. 
\begin{corollary}\label{cor:ZUSpure}
	A  bipartite pure state $|\Psi\>$ in $\caH_\rmA\otimes \caH_\rmB$ is a ZUS with respect to $\scrB$ iff  the reduced state $\rho_\rmA$ is a weighted sum of component projectors in $\scrP(\scrB)$. 
\end{corollary}
\Crref{cor:ZUSpure} is a special case of  \thref{thm:ZUS2} and also follows from \lref{lem:PerfectDist} and Supplementary Lemma~2. Here $\rho_\rmA$ is a weighted sum of component projectors  iff $\rho_\rmA$ is diagonal with respect to each basis $\caB$ in $\scrB$ (cf. Supplementary note C). 
As an implication of \crref{cor:ZUSpure}, the reduced state $\rho_\rmA$ of any (pure or mixed) ZUS $\rho$ with respect to $\scrB$ is a weighted sum of component projectors in $\scrP(\scrB)$, given that any ZUS is a convex mixture of pure ZUSs. 
The equivalent classes of pure ZUSs are determined in the Methods section.

\bigskip

\noindent\textbf{Applications to quantum certification and verification}\\
Our results on ZUSs have immediate implications for the verification of MESs.  Suppose Alice and Bob want to create the  MES $|\Phi\>=\sum_{j=0}^{d_\rmA-1}|jj\>/\sqrt{d_\rmA}$ in this way:  Bob first creates a MES in his lab and then sends one particle of the entangled pair to Alice via a quantum channel. 
To verify the resulting state $\rho$, they can perform tests based on  correlated local projective measurements such that only the target state can pass all tests with certainty \cite{HayaMT06,PallLM18,ZhuH19O,ZhuH19AdS,ZhuH19AdL}. Suppose Alice can perform projective measurements from the set $\scrB=\{\caB_x\}_{x=1}^m$ in which $\caB_x$ is chosen with probability $\mu_x>0$. For each choice $\caB_x$, she asks Bob to perform the measurement on the 
  conjugate basis $\caB_x^*$ and return the outcome. The test is passed if Bob and Alice obtain the same outcome \cite{ZhuH19O}.

If Bob is honest, then  the average probability that $\rho$ passes each test is $\tr(\rho\Omega)$, where 
$\Omega=\sum_{x=1}^m \mu_x P(\caB_x)$ \cite{ZhuH19O}  is known as the verification operator  and
\begin{equation}
P(\caB_x):=\sum_{|\psi\> \in \caB_x} |\psi\>\<\psi|\otimes |\psi^*\>\<\psi^*|
\end{equation}
is a test projector.
Note that $|\Phi\>$ is an eigenstate of $P(\caB_x)$ and  $\Omega$ with eigenvalue 1 and so can pass each test with certainty. In addition, $|\Phi\>$ can be reliably verified by this protocol iff the maximum eigenvalue of $\Omega$ is nondegenerate, that is, the pass eigenspace of $\Omega$ has dimension 1 \cite{PallLM18,ZhuH19O,ZhuH19AdS,ZhuH19AdL}. This is the case iff the basis set $\scrB$ is irreducible by the following theorem proved in the Supplementary Information.

\begin{theorem}\label{thm:degeneracy}
	Suppose $\scrB=\{\caB_x\}_{x=1}^m$ is a set of orthonormal bases in $\caH_\rmA$, and
	 $\Omega=\sum_{x=1}^m \mu_x P(\caB_x)$ with $\mu_x>0$ and $\sum_{x=1}^m \mu_x=1$. Then the degeneracy of the maximum eigenvalue 1 of $\Omega$  equals the number of connected components of the transition graph $G(\scrB)$. 
	 This eigenvalue is nondegenerate iff the basis set $\scrB$ is irreducible.
\end{theorem}

Next, suppose Bob is not honest. Then  Alice cannot distinguish states that are equivalent under local operations of Bob. Nevertheless, she can still verify the MES $|\Phi\>$ up to equivalence. Thanks to \thref{thm:ZUS}, the uncertainty game described before actually provides a verification protocol whenever the basis set $\scrB$ of Alice is irreducible. Note that Bob can pass each test (guess each measurement outcome) of Alice with certainty only if the state $\rho$ prepared is a MES.
Surprisingly, the requirement on the measurement bases of  Alice  remains 
the same when Bob becomes dishonest.  In addition, the measurements of Bob required to attain the maximum guessing probability are essentially uniquely determined by the counterpart of Alice. These results are of interest to semi-device-independent quantum certification and verification 
\cite{MayeY04,SupiH16,MccuPBM16,SupiB20,EiseHWR20}

\bigskip
\noindent\textbf{Implications for quantum steering}\\
Our study also has implications for Einstein-Podolsky-Rosen steering or quantum steering \cite{EinsPR35,WiseJD07,SupiH16,CavaS17,UolaCNG20}, which is clear if we interchange the measurement order in the above verification protocol.  In each test Alice asks Bob to perform the measurement on the basis $\caB_x^*$ with probability $\mu_x>0$ for $x=1,2,\ldots, m$ and return the  outcome. Then Alice performs the projective measurement on  $\caB_x\in \scrB$, and the test is passed if she obtains the same outcome as Bob. Alternatively, Alice can choose the two-outcome POVM $\{|\psi_{xk}\>\<\psi_{xk}|, 1_\rmA-|\psi_{xk}\>\<\psi_{xk}|\}$ if Bob obtains outcome $k$. Suppose Alice and Bob share the state $\rho$ and Bob  performs the POVM $\{\Pi_{xk}\}_k$ instead of the projective measurement on $\caB_x^*$. Then the probability that Bob  passes the test reads 
\begin{equation}
\sum_k \<\psi_{x k}|\sigma_{xk}|\psi_{x k}\>,\quad \sigma_{xk}=\tr_\rmB[\rho (1_\rmA\otimes \Pi_{xk})],
\end{equation}
which is a variant of \eref{eq:GuessProb}. The subnormalized  states $\sigma_{xk}$  satisfy the normalization condition $\sum_k\sigma_{xk}=\rho_\rmA$ and form an \emph{ensemble} of $\rho_\rmA$ for each $x$. The collection of ensembles $\{\{\sigma_{xk}\}_k\}_x$ is known as an \emph{assemblage} \cite{Puse13,CavaS17,UolaCNG20}. 
If Bob is honest, then the assemblage generated is $\{\{|\psi_{xk}\>\<\psi_{xk}|/d_\rmA\}_k\}_x$.
When $\scrB$ is irreducible, it turns out only this  assemblage can pass each test with certainty, as shown in the following lemma and proved in the Supplementary Information.  
\begin{lemma}\label{lem:AssemblageTest}
Suppose $\scrB=\{\caB_x\}_{x=1}^m$ with $\caB_x=\{|\psi_{xk}\>\}_k$ is an irreducible set of orthonormal bases in $\caH_\rmA$. Suppose $\{\{\sigma_{xk}\}_k\}_x$	is an assemblage for $\rho_\rmA$  that satisfies	$\sum_k \<\psi_{x k}|\sigma_{xk}|\psi_{x k}\>=1$ for each $x$. Then $\rho_\rmA$ is completely mixed, and $\sigma_{xk}=|\psi_{x k}\>\<\psi_{x k}|/d_\rmA$ for each $x$ and $k$. 
\end{lemma}
Here the condition $\sum_k \<\psi_{x k}|\sigma_{xk}|\psi_{x k}\>=1$  means the assemblage $\{\{\sigma_{xk}\}_k\}_x$ can pass each test of Alice with certainty. By virtue of \thref{thm:ZUS}, we can further show that the assemblage $\{\{|\psi_{xk}\>\<\psi_{xk}|/d_\rmA\}_k\}_x$ can only be generated by a MES, as stated in the following theorem and proved in the Supplementary Information.
\begin{theorem}\label{thm:MESassemblage}
Given the basis set $\scrB$ in \lref{lem:AssemblageTest}, 
suppose $\rho$ is a bipartite state on $\caH_\rmA\otimes\caH_\rmB$ that can generate the assemblage $\{\{|\psi_{xk}\>\<\psi_{xk}|/d_\rmA\}_k\}_x$ under the measurements of Bob. Then  $d_\rmB\geq d_\rmA$, and $\rho$ is a MES and a ZUS with respect to $\scrB$.
\end{theorem}
Thanks to \lref{lem:AssemblageTest} and \thref{thm:MESassemblage}, the tests of Alice can verify the assemblage $\{\{|\psi_{xk}\>\<\psi_{xk}|/d_\rmA\}_k\}_x$,
which in turn can verify the MES whenever $\scrB$ is irreducible. In addition, \Thref{thm:MESassemblage} offers a general recipe for constructing assemblages that are characteristic of MESs.
These results establish intimate connections between  uncertainty relations, quantum entanglement, and quantum steering, which are of intrinsic interest  to foundational studies. Meanwhile, these results are instructive to studying remote state preparation \cite{BennDSS01,BerrS03} and semi-device-independent self testing \cite{MayeY04,SupiH16,MccuPBM16,SupiB20,EiseHWR20}.

\section{Discussion} 
Zero uncertainty states in the presence of a quantum memory are particular quantum states that can eliminate uncertainties of incompatible von Neumann observables once assisted by suitable measurements on the memory. In this work we determined all ZUSs with respect to any given set of nondegenerate observables in the presence of a quantum memory. To achieve this goal we introduced several useful tools that apply to an arbitrary set of observables, in sharp contrast with most previous approaches, which  only apply to two observables or complementary observables.
In addition, we  determined the minimum entanglement required to construct a ZUS.
Our study shows that  all ZUSs are MESs for a generic set of two or more observables even if these observables are only weakly incompatible.
In this way we establish a simple and precise connection between ZUSs and MESs. 
This connection may shed light on the uncertainty principle in the presence of a quantum memory. It is of intrinsic interest  to studying a number of fascinating topics, including the uncertainty principle, quantum entanglement, and quantum  steering. Moreover, it has direct applications in  semi-device-independent quantum certification and verification, which is currently an active research field.

\bigskip

\noindent\textbf{Methods}\\
\noindent\textbf{Zero uncertainty states with least entanglement}\\
Here we determine the minimum entanglement required to construct a ZUS, following the premises and notations in \thref{thm:ZUS2}.

Given the basis set $\scrB$ in \thref{thm:ZUS2}, define $\Lambda_\scrB$ as the CPTP map acting on quantum states on $\caH_\rmA\otimes \caH_\rmB$ that  removes coherence between different component subspaces, that is,
\begin{equation}\label{eq:PhiB}
\Lambda_\scrB(\rho)=\sum_a (P_a\otimes 1_\rmB)\rho(P_a\otimes 1_\rmB)=\sum_a \rho_a=\bigoplus_a \rho_a,
\end{equation}
where $\rho_a=(P_a\otimes 1_\rmB)\rho(P_a\otimes 1_\rmB)$ and $P_a$ are component projectors defined in \eref{eq:ComponentProj}.  Note that the map $\Lambda_\scrB$ can be realized by LOCC. 
In addition, $\rho$ is a ZUS with respect to $\scrB$ iff $\rho_\scrB:=\Lambda_\scrB(\rho)$ is. When $\rho$ is a ZUS, $\rho_\scrB$ is a direct sum of subnormalized MESs  $\rho_a$ according to \thref{thm:ZUS2}.

The  \emph{component vector} is defined as
\begin{align}
\bfq(\rho,\scrB):=(q_a)_a,\quad q_a:=\tr[\rho (P_a\otimes 1_\rmB)]=\tr(\rho_a);
\end{align}
it is invariant under local operations of Bob and is very useful to studying the entanglement properties and equivalent classes of  ZUSs. Note that 
 $\rho_\scrB$ and $\rho$ share the  same component vector. The following theorem is proved in  the Supplementary Information.
\begin{theorem}\label{thm:ZUSEB}
Suppose $E$ is an entanglement measure, then any ZUS $\rho$ with component vector $\bfq(\rho,\scrB)=(q_a)_a$
satisfies 
\begin{align}
\!\!E(\rho)\geq E(\rho_\scrB)\geq \sum_a  q_a E(|\Phi(r_a)\rangle)\geq E(|\Phi(r_{\min})\rangle), \label{eq:ZUSEB}
\end{align}
where $|\Phi(r_a)\rangle$  is a MES of Schmidt rank $r_a$. The second inequality in \eref{eq:ZUSEB} is saturated if $E$ is a convex entanglement measure. 
\end{theorem}
Here  $r_a=\tr(P_a)$ is the component rank defined in \eref{eq:ComponentProj}, and $r_{\min}=\min_{1\leq a\leq g} r_a$ is the minimum  component rank.
\Thref{thm:ZUSEB} applies to any entanglement measure $E$ that is monotonic under selective and nonselective LOCC. In addition, the lower bounds for $E(\rho)$ only depend on the values of the measure $E$ at pure MESs.

 When $\rho_\scrB\neq \rho$,  the inequality $E(\rho)\geq E(\rho_\scrB)$ in \eref{eq:ZUSEB} is strict for many entanglement measures, including the entanglement of formation, as shown in 
\crref{cor:ZUSEF1} below. To determine least entangled  ZUSs,  we can assume the condition $\rho=\rho_\scrB$, so  $\rho$ has no coherence between different component subspaces.
Such  a ZUS is called \emph{economical}.
In addition, the third inequality in \eref{eq:ZUSEB} is usually strict unless $q_a=0$ when  $r_a>r_{\min}$ (cf.~\crref{cor:ZUSEF2} below). An economical ZUS $\rho$  with $q_a=0$ for all  $r_a>r_{\min}$ is called a \emph{ZUS with least entanglement} (ZUSLE) since it can saturate the ultimate lower bound in \eref{eq:ZUSEB} for every convex entanglement measure. 
Such a state can be expressed as follows,
\begin{equation}\label{eq:ZUSLE}
\rho=\rho_\scrB=\sum_{a|r_a=r_{\min}} \rho_a=\bigoplus_{a|r_a=r_{\min}} \rho_a.
\end{equation}
It has no coherence between different component subspaces, and its local support for Alice can only contain component subspaces with the minimum component rank $r_{\min}$. The operational significances of 
 ZUSLEs can be summarized as follows.
\begin{corollary}\label{cor:ZUSLE1}
	Every ZUS on $\caH_\rmA\otimes \caH_\rmB$ with respect to $\scrB$  can be turned into a ZUSLE by LOCC.  In addition, all ZUSLEs can be turned into each other by LOCC.
\end{corollary}
\begin{corollary}\label{cor:ZUSLE2}
	A ZUS on $\caH_\rmA\otimes \caH_\rmB$ with respect to $\scrB$ is a ZUSLE iff it can be created from $|\Phi(r_{\min})\rangle$ by LOCC.
\end{corollary}
\Crref{cor:ZUSLE1} follows from \thref{thm:ZUS2} and \crref{cor:MES}. 
\Crref{cor:ZUSLE2} follows from \Crref{cor:ZUSLE1} and \crref{cor:ZUSEF2} below.
This operational characterization of ZUSLEs is independent of specific entanglement measures, which complements the operational definition of ZUSs in the main text.

\begin{corollary}\label{cor:ZUSEF1}
	Any ZUS $\rho$ with component vector $\bfq(\rho,\scrB)=(q_a)_a$
	satisfies $E_\rmF(\rho)\geq\sum_a q_a\log_2 r_a$. The lower bound is saturated iff  $\rho=\rho_\scrB$. 
\end{corollary}
\begin{corollary}\label{cor:ZUSEF2}
	Any ZUS $\rho$ with respect to $\scrB$ satisfies  $E_\rmF(\rho)\geq \log_2 r_{\min}$, which is saturated iff  $\rho$ is a ZUSLE. 
\end{corollary}
\Crref{cor:ZUSEF1} is proved in the Supplementary Information. 
\Crref{cor:ZUSEF2} follows from  \crref{cor:ZUSEF1} (cf. \thref{thm:ZUSEB}). When $\scrB$ is irreducible, the bound $E_\rmF(\rho)\geq\sum_a q_a\log_2 r_a$ reduces to  $E_\rmF(\rho)\geq\log_2 d_\rmA$, which is expected in view of \thref{thm:ZUS} and \lref{lem:MES}. In general, the lower bound may be seen as a weighted average of bounds associated with individual irreducible components of $\scrB$.  
Incidentally, the bounds in \crsref{cor:ZUSEF1} and \ref{cor:ZUSEF2} still hold if $E_\rmF$ is replaced by any  entanglement measure that coincides with $E_\rmF$ on pure states, such as 
the relative entropy of entanglement \cite{HoroHHH09}.

\bigskip
\noindent\textbf{Equivalent classes of zero uncertainty states}\\
Here we clarify the  equivalent classes of ZUSs under local operations of Bob. 
Suppose $\rho$ is a ZUS with respect to $\scrB$ and has component vector $\bfq(\rho,\scrB)=(q_a)_a$. According to \crref{cor:ZUSpure}, (cf. \thref{thm:ZUS2}, \lref{lem:MES}, and  \crref{cor:MES}), $\rho_\rmA$ is a weighted sum of component projectors, 
\begin{equation}
\rho_\rmA=\sum_a \frac{q_aP_a}{r_a}.
\end{equation}
Two ZUSs have the same reduced state  and thus same measurement statistics for Alice iff they have the same component vector. So  the equivalent classes of pure ZUSs are completely characterized by  component vectors. 
\begin{corollary}\label{cor:ZUSpureEqui}
Two pure ZUSs with respect to  $\scrB$ are equivalent iff they have the same component vector. 	
\end{corollary}

Given a ZUS $\rho$ with respect to the basis set $\scrB$, denote by  $\caH_{\rmB,a}$ the support of  $\tr_\rmA(\rho_a)$ 
and  $Q_a$  the corresponding projector; then
 $Q_a$ and $Q_b$ are orthogonal whenever $a\neq b$  by \thref{thm:ZUS2}. In addition,
 we have
\begin{align}
\rho_a&=(1_\rmA\otimes Q_a)\rho(1_\rmA\otimes Q_a),\\
\rho_\scrB&=\sum_a (1_\rmA\otimes Q_a)\rho(1_\rmA\otimes Q_a).
\end{align}
So  $\rho$ can be turned into $\rho_\scrB$ by local operations of Bob. Thanks to \crref{cor:MES}, $\rho_\scrB$ can  further be turned into a direct sum of pure MESs of the form $\bigoplus_a q_a |\Phi_a\>\<\Phi_a|$,
where $|\Phi_a\>$ is a normalized  MES in $\caH_{\rmA,a}\otimes \caH_{\rmB,a}$ (a product state when $r_a=1$). These observations lead to the following corollary, which complements \crref{cor:ZUSpureEqui}. 
\begin{corollary}\label{cor:ZUSeco}
	Every ZUS on $\caH_\rmA\otimes \caH_\rmB$ with respect to $\scrB$ can be turned into an economical ZUS with the same component vector by local operations of Bob. Two economical ZUSs are equivalent iff they have the same component vector. 
\end{corollary}
Thanks to \crref{cor:ZUSeco}, equivalent classes of economical ZUSs with respect to $\scrB$ are in one-to-one correspondence with component vectors, which form a probability simplex of dimension $g-1$, where $g$ is the number of irreducible  components of  $\scrB$. 
In particular, two ZUSLEs are equivalent iff they have the same component vector. 
If there is only one component subspace of dimension $r_{\min}$, then all ZUSLEs are equivalent.

\bigskip
\noindent\textbf{ACKNOWLEDGEMENTS}\\
This work is  supported by   the National Natural Science Foundation of China (Grant No.~11875110) and  Shanghai Municipal Science and Technology Major Project (Grant No.~2019SHZDZX01). \\

\bibliographystyle{naturemag}
\bibliography{all_references}

\bigskip


\clearpage


\setcounter{equation}{0}
\setcounter{figure}{0}
\setcounter{table}{0}
\setcounter{theorem}{0}
\setcounter{lemma}{0}
\setcounter{remark}{0}
\setcounter{proposition}{0}
\setcounter{corollary}{0}
\setcounter{section}{0}



 \onecolumngrid
\begin{center}
	\textbf{\large Zero Uncertainty States in the Presence of  Quantum Memory:\\ Supplementary Information}
\end{center}
\twocolumngrid

In this Supplementary Information, we prove the key results presented in the main text, including \thsref{thm:ZUS} and \ref{thm:degeneracy}-\ref{thm:ZUSEB}. Several auxiliary results, including \lsref{lem:MES}-\ref{lem:AssemblageTest} and \crref{cor:ZUSEF1}, are also proved for completeness. 

\subsection*{Supplementary note A: Proof of  of \lref{lem:MES}}
In this section we prove \lref{lem:MES} in the main text, which clarifies the structures and properties of maximally entangled states (MESs); cf.~\rcite{LiZFF12}. This lemma implies the existence of mixed states that are maximally entangled when $d_\rmB\geq 2d_\rmA$, although these states are mixed in a trivial way. Incidentally, for a given spectrum, there is a different definition of maximally entangled mixed states---mixed states whose entanglement cannot be increased by global unitary transformations \cite{VersAM01}. However, this definition usually depends on a specific entanglement measure, although entanglement of formation, negativity, and relative entropy of entanglement lead to the same  states in the case of two-qubits \cite{VersAM01}. In addition, it is extremely difficult to determine such maximally entangled mixed states in general  if not impossible. As far as we know, the problem has not been solved yet even for two qutrits. By contrast, the definition of MESs (pure or mixed) that we consider is universal because it builds on transformations under LOCC and  is independent of a specific entanglement measure. In addition, their structures have a simple description for all bipartite systems. These merits are quite appealing to establishing a universal connection between maximum entanglement and zero uncertainty.

\begin{proof}[Proof of \lref{lem:MES}] 
	First, note that any pure state in $\caH_\rmA\otimes \caH_\rmB$ can be created from 	
	\begin{equation}
	|\Phi\>=\frac{1}{\sqrt{d_\rmA}}\sum_{j=0}^{d_\rmA-1}|jj\>
	\end{equation}	
	under LOCC according to the majorization criterion \cite{Niel99}, given that $d_\rmB\geq d_\rmA$. Since any mixed state is a convex mixture of pure states, it follows that any state on  $\caH_\rmA\otimes \caH_\rmB$
	can be created from  $|\Phi\rangle$
	under LOCC. Therefore, $|\Phi\rangle$ is indeed a MES according to the concrete definition presented in the main text; it is referred to as the canonical MES. In addition, all pure MESs on $\caH_\rmA\otimes \caH_\rmB$ are equivalent to $|\Phi\rangle$
	under local unitary transformations of Bob.

	Any state $\rho$ on $\caH_\rmA\otimes \caH_\rmB$	satisfies the following inequalities:
	\begin{align}\label{seq:EntropyInq}
	-\log_2 d_\rmA&\leq -H(\rmA|\rmB)_\rho\leq E_\rmR(\rho)\leq E_\rmF(\rho)\leq S(\rho_\rmA)\nonumber\\
	&\leq  \log_2 d_\rmA. 
	\end{align}
	Here the first inequality can be derived as follows,
	\begin{align}
	H(\rmA|\rmB)_\rho&=S(\rho)-S(\rho_\rmB)\leq S(\rho_\rmA)\leq \log_2 d_\rmA. 
	\end{align}
	The second inequality in Supplementary \eref{seq:EntropyInq}  is derived in \rcite{PlenVP00} (cf.~\rcite{ZhuHC17C}). The third inequality follows from the three  facts:  $E_\rmR(\rho)$ and $E_\rmF(\rho)$ coincide on pure states; $E_\rmF(\rho)$ is an entanglement measure based on the convex roof; $E_\rmR(\rho)$ is convex in $\rho$ \cite{HoroHHH09}.  The fourth inequality follows from the convex-roof  definition of $E_\rmF(\rho)$ and the concavity of the von Neumann entropy. The last inequality  is well known. 
	
	If $\rho$ is a MES, then we can create the MES $|\Phi\>$ from $\rho$ using LOCC, so $E_\rmR(\rho)\geq E_\rmR(\Phi)=\log_2 d_\rmA$, which together with Supplementary \eref{seq:EntropyInq} implies the equality $E_\rmR(\rho)=\log_2 d_\rmA$  and confirms the implication  $1\Rightarrow 3$. The implications  $2\Rightarrow 3$ and  $3\Rightarrow 4$ also follow from Supplementary \eref{seq:EntropyInq}. If statement 5 holds, then $\rho$ can be transformed into $|\Phi\>$ under LOCC (local operations of Bob alone are sufficient), so $\rho$ is a MES. In addition,
	the equality $H(\rmA|\rmB)_\rho=-\log_2 d_\rmA$ can be verified by straightforward calculation. Therefore, statement~5 implies statements 1 and 2. To prove \lref{lem:MES} in the main text, it remains to prove the implication $4\Rightarrow 5$.

	If $E_\rmF(\rho)=\log_2 d_\rmA$, then all pure states in the support of $\rho$ have the same entanglement of formation and are MESs.
	Let  $\rho=\sum_s \lambda_s |\Psi_s\>\<\Psi_s|$ be a spectral decomposition; then each $|\Psi_s\>$ is a MES  and can be expressed as follows,
	\begin{equation}
	|\Psi_s\>=\frac{1}{\sqrt{d_\rmA}}\sum_j |j\>\otimes |\varphi_{sj}\>,
	\end{equation}
	where the kets $|\varphi_{sj}\>$ for a given $s$ are orthonormal. In addition,  for a given pair of  $s$ and $t$ with $s\neq t$, the ket  $(|\Psi_s\>+|\Psi_t\>)/\sqrt{2}$  is  maximally entangled, so  the kets $(|\varphi_{sj}\>+|\varphi_{tj}\>)/\sqrt{2}$ are orthonormal, which implies that
	\begin{equation}
	\<\varphi_{sj}|\varphi_{tk}\>+\<\varphi_{tj}|\varphi_{sk}\>=0\quad \forall j,k=0,1,\ldots, d_\rmA-1. 
	\end{equation}
	Similarly, $(|\Psi_s\>+\rmi|\Psi_t\>)/\sqrt{2}$ is  maximally entangled, which implies that 
	\begin{equation}
	\<\varphi_{sj}|\varphi_{tk}\>-\<\varphi_{tj}|\varphi_{sk}\>=0\quad \forall j,k=0,1,\ldots, d_\rmA-1. 
	\end{equation}
	As an implication of the above two equations, we have $\<\varphi_{sj}|\varphi_{tk}\>=0$ for all $j,k$, so $\tr_\rmA(|\Psi_s\>\<\Psi_s|)$ and $\tr_\rmA(|\Psi_t\>\<\Psi_t|)$ have orthogonal supports whenever $s\neq t$. Therefore, every spectral decomposition of $\rho$ has the properties described in statement~5, which confirms the implication $4\Rightarrow 5$.
\end{proof}

\subsection*{Supplementary note B: Proof of \lref{lem:PerfectDist}}
In this section we prove \lref{lem:PerfectDist} in the main text, which is crucial to understanding ZUSs in the presence of a quantum memory and to proving the main result \thref{thm:ZUS}. If $\rho$ is a MES of Schmidt rank $d_\rmA$, then $\rho_\rmA$ is completely mixed and is thus diagonal with respect to any orthonormal basis, so  the ensemble $\scrS(\rho, \caB)$ of reduced states of Bob defined in \eref{eq:ensemble} in the main text  is perfectly distinguishable for any projective measurement   $\caB$ of Alice as expected. 

\Lref{lem:PerfectDist} in the main text is an immediate corollary of the following lemma, which is of some independent interest.
\begin{lemma}
	Suppose $\{|\varphi_j\>\}_j$ is an orthonormal basis in $\caH_\rmA$ and $M$ is a linear operator from  $\caH_\rmA$ to $\caH_\rmB$. Then the vectors in the ensemble $\{M|\varphi_j\>\}_j$ are pairwise mutually orthogonal iff $\{|\varphi_j\>\}_j$ is an eigenbasis of $M^\dag M$. 
\end{lemma}
Here we take the convention that a zero vector is orthogonal to all  vectors. If $d_\rmB=d_\rmA$ and $M$ is an invertible  operator, then $\{M|\varphi_j\>\}_j$ is an orthogonal basis in $\caH_\rmB$ iff $\{|\varphi_j\>\}_j$ is an eigenbasis of $M^\dag M$.

\begin{proof}
	If $\{|\varphi_j\>\}_j$ forms an eigenbasis of $M^\dag M$, then $M^\dag M|\varphi_j\>\propto |\varphi_j\>$, so that  $M|\varphi_j\>$ are mutually orthogonal. Conversely, if  $M|\varphi_j\>$ are mutually orthogonal, then $M^\dag M|\varphi_j\>$ for each $j$ is orthogonal to $|\varphi_k\>$ for all $k\neq j$. Therefore,  $M^\dag M|\varphi_j\>\propto |\varphi_j\>$, which means $\{|\varphi_j\>\}_j$ forms an eigenbasis of $M^\dag M$. 
\end{proof}

\subsection*{Supplementary note C: Operators that are diagonal with respect to a basis set}
An operator is normal if it  commutes with its hermitian conjugate \cite{Meye00}. 
It is well known that a normal operator  can be diagonalized in a suitable orthonormal basis, and vice versa. 
Here we determine those  operators that are diagonal with respect to each basis in a set of orthonormal bases. It turns out that such  operators are determined by component projectors associated with  the basis set as introduced in the main text. Accordingly, the dimension of the space composed of these  operators is equal to the number of connected components of the transition graph of the basis set. 
This result is crucial to establishing the connection between ZUSs and MESs, as revealed in   \thref{thm:ZUS}.
\begin{lemma}\label{slem:DiagonalOperator}
	Let $\scrB=\{\caB_x\}_{x=1}^m$ be a set of orthonormal bases in $\caH_\rmA$ and $M$  an operator on $\caH_\rmA$. Then $M$ is diagonal with respect to each basis in $\scrB$ iff $M$ is a linear combination of component projectors in $\scrP(\scrB)$ as defined in \eref{eq:ComponentProj} in the main text. 
\end{lemma}
\begin{remark}
	$M$ is diagonal with respect to the basis $\caB_x$ iff $M$ commutes with all projectors onto basis states in $\caB_x$.
	Such an operator is necessarily normal. Supplementary \lref{slem:DiagonalOperator} implies that the commutant of the operator set $\cup_{x=1}^m\{|\psi\>\<\psi|: |\psi\>\in \caB_x\}$ is generated by component projectors. When $\scrB$ is irreducible, there is a unique component projector, which coincides with the identity operator on $\caH_\rmA$, so $M$ is proportional to the identity operator. These results in particular apply to density operators. 
\end{remark}

\begin{proof}
	If $M$ is diagonal with respect to each basis in $\scrB$, then $M$ is normal and  each basis state in each basis $\caB_x$ in $\scrB$ is an eigenstate of $M$. If two states are not orthogonal, then the eigenvalues are necessarily the same given that eigenstates associated with different eigenvalues of a normal operator are orthogonal \cite{Meye00}.  So all  states corresponding to the vertices in a connected component of the transition graph $G(\scrB)$ share a same eigenvalue. Therefore, $M$ is a linear combination of component projectors in $\scrP(\scrB)$, in which case $M$ is indeed diagonal with respect to  each basis in  $\scrB$. 
\end{proof}

\subsection*{Supplementary note D: Proof of \thref{thm:ZUS}}

\begin{proof}
	First, suppose $\rho$ is a pure state. If $d_\rmB\geq d_\rmA$ and $\rho$ is a MES,  then the states in the ensemble $\scrS(\rho, \caB)$ defined in \eref{eq:ensemble} in the main text are mutually orthogonal and thus perfectly distinguishable for any orthonormal basis $\caB$ in $\caH_\rmA$ (cf. \lref{lem:PerfectDist} in the main text).  So $\rho$ is a ZUS with respect to $\scrB$. 
	
	Conversely,  if $\rho$ is a ZUS with respect to $\scrB$, then the states in the ensemble $\scrS(\rho, \caB)$ for each basis $\caB\in \scrB$  are perfectly distinguishable.
	So $\rho_\rmA$ is diagonal with respect to each basis $\caB$ in $\scrB$ by \lref{lem:PerfectDist} in the main text. Since the basis set $\scrB$ is irreducible, $\rho_\rmA$ must be a completely mixed state by Supplementary \lref{slem:DiagonalOperator}, which implies that $d_\rmB\geq d_\rmA$ and that $\rho$ is a MES. 
	
	Next, suppose $\rho$ is mixed. If $d_\rmB\geq d_\rmA$ and  $\rho$ is a MES, then $\rho$  has a spectral decomposition
	$\rho=\sum_s \lambda_s |\Psi_s\>\<\Psi_s|$ in which  each $|\Psi_s\>$ is a MES  by \lref{lem:MES} in the main text and is thus a ZUS with respect to $\scrB$. In addition, the reduced states $\tr_\rmA(|\Psi_s\>\<\Psi_s|)$ have mutually orthogonal supports, so $\rho$ is also   a ZUS by \pref{pro:ZUSsum} in the main text. Alternatively, this conclusion follows from \crref{cor:MES} and \pref{pro:ZUSCPTP} in the main text given the above conclusion on pure states.

	Conversely, if $\rho$ is a ZUS, then every pure state in its  support is a ZUS by \pref{pro:ZUSsupport} in the main text and thus a MES given the above discussion; in addition,  $d_\rmB\geq d_\rmA$. Therefore, $E_\rmF(\rho)=\log_2 d_\rmA$, so that $\rho$ is a MES by \lref{lem:MES} in the main text. 	
\end{proof}

\subsection*{Supplementary note E: Optimal measurements of Bob}
Here we determine the optimal measurement of Bob required to maximize the guessing probability. For any given ZUS $\rho$, it turns out the optimal measurement of Bob on the support of $\rho_\rmB$ is uniquely determined by the counterpart of Alice.

When $\rho$ is a ZUS and thus a MES, to determine the optimal measurement of Bob, note that $\rho$ can be expressed as a tensor product of a pure MES and an ancillary state by \lref{lem:MES} in the main text. Without loss of generality, we may assume that the support of $\rho_\rmB$ coincides with $\caH_\rmB$ since   modification of POVM elements outside this support does not affect the guessing probability. 
Then $\caH_\rmB$  has  a  decomposition $\caH_\rmB=\caH_{\rmB_1}\otimes \caH_{\rmB_2}$ with $\dim(\caH_{\rmB_1})=d_\rmA$ 
such that $\rho=|\Phi'\>\<\Phi'|\otimes \tau$, where $|\Phi'\>$ is a pure MES in $\caH_\rmA\otimes \caH_{\rmB_1}$, and $\tau$ is a full-rank density operator on $\caH_{\rmB_2}$.

If Alice performs the projective measurement on the basis $\caB_x=\{|\psi_{xk}\>\}_k$ and obtains outcome $k$, then the normalized reduced state of Bob reads $\rho_{xk}'\otimes \tau$,  where  
\begin{equation}
\rho_{xk}'=d_\rmA \<\psi_{xk}|\Phi'\>\<\Phi'|\psi_{xk}\>.
\end{equation}
When $|\Phi'\>=|\Phi\>$  is the canonical MES for example, we have $\rho_{xk}'=|\psi_{xk}^*\>\<\psi_{xk}^*|$,  where $|\psi_{xk}^*\>$ denotes the complex conjugate of $|\psi_{xk}\>$ in the computational basis. For any given basis $\caB_x$, note  that $\rho_{xk}'$ are  mutually  orthogonal rank-1 projectors and satisfy $\sum_k\rho_{xk}'=1_{\rmB_1}$. 
To attain the maximum guessing probability 1, the POVM $\{\Pi_{xk}\}_k$ of Bob must satisfy the condition $\Pi_{xk}\geq \rho_{xk}'\otimes 1_{\rmB_2}$ for all $k$.  This result implies that 
$\Pi_{xk}= \rho_{xk}'\otimes 1_{\rmB_2}$
in view of   the normalization condition 
\begin{equation}
\sum_k \Pi_{xk}=1_\rmB=1_{\rmB_1}\otimes 1_{\rmB_2}.
\end{equation}
Therefore,  given any ZUS $\rho$, the optimal measurement of Bob on the support of $\rho_\rmB$ is uniquely determined by the counterpart of Alice.

\subsection*{Supplementary note F: Proofs of \thsref{thm:degeneracy}, \ref{thm:MESassemblage} and \lref{lem:AssemblageTest}}

\begin{proof}[Proof of \thref{thm:degeneracy}]
	Suppose the orthonormal basis $\caB_x$ consists of the kets $|\psi_{x k}\>$	for $k=0,1,\ldots, d_\rmA-1$. Define subnormalized vectors
	\begin{equation}
	|\tilde{v}_{x k}\>:=\sqrt{\mu_x}|\psi_{x k}\>\otimes |\psi_{x k}^*\>;
	\end{equation}	
	then we have $\Omega=\sum_{x k} |\tilde{v}_{x k}\>\<\tilde{v}_{x k}|$. Let $M$ be the Gram matrix of the set of vectors $|\tilde{v}_{x k}\>$ for $x=1,2,\ldots, m$ and $k=0,1,\ldots, d_\rmA-1$, that is, 
	\begin{equation}
	M_{xk,yl}=\<\tilde{v}_{xk}|\tilde{v}_{yl}\>=\sqrt{\mu_x\mu_y}|\<\psi_{xk}|\psi_{yl}\>|^2.
	\end{equation}
	Then $\Omega$ and $M$ have the same nonzero eigenvalues, including degeneracies. Note that $M$ is a positive semidefinite matrix whose entries are nonnegative. In addition, the adjacency matrix of the transition graph $G(\scrB)$ can be constructed from $M$ by replacing nonzero off-diagonal entries with the constant~1.

	If the transition graph $G(\scrB)$  has $g$  connected components, then $M$ decomposes into a direct sum of $g$ positive semidefinite irreducible matrices, which are in one-to-one correspondence with the connected components of $G(\scrB)$. Recall that  a nonnegative matrix is irreducible if it has no nontrivial invariant coordinate subspace. Let $M^{(a)}$ be the irreducible matrix associated with the irreducible component $G_a$. Define the vector $\bm{u}^{(a)}$ with entries
	\begin{equation}
	u^{(a)}_{xk}=\sqrt{\mu_x} \quad \mbox{for } |\psi_{xk}\>\in \caB_x^a.
	\end{equation}
	Then $\bm{u}^{(a)}$ is an  eigenvector of  $M^{(a)}$ with eigenvalue 1. Moreover, according to the Perron-Frobenius theorem (see Chap.~8 of \rcite{Meye00} for example),   the maximum  eigenvalue of $M^{(a)}$ is equal to 1 and is nondegenerate. Therefore, the maximum eigenvalue of $M$ is $g$-fold degenerate, and the same holds for $\Omega$. In particular, the maximum eigenvalue of $\Omega$ is nondegenerate iff  $G(\scrB)$  is connected, in which case the basis set $\scrB=\{\caB_x\}_{x=1}^m$ is irreducible. 
\end{proof}

\begin{proof}[Proof of \lref{lem:AssemblageTest}] For each basis $\caB_x$, the equality  $\sum_k \<\psi_{x k}|\sigma_{xk}|\psi_{x k}\>=1$  implies that $\sigma_{xk}\propto |\psi_{xk}\>\<\psi_{xk}|$ for each $k$ and that  $\rho_\rmA$ is diagonal with respect to $\caB_x$. According to Supplementary \lref{slem:DiagonalOperator}, $\rho_\rmA$ is necessarily completely mixed given that the basis set  $\scrB$ is irreducible. Now, for each $x$, the requirement $\sum_k\sigma_{xk}=\rho_\rmA$  implies that $\sigma_{xk}=|\psi_{x k}\>\<\psi_{x k}|/d_\rmA$ for each $k$, so the assemblage $\{\{\sigma_{xk}\}_k\}_x$ is identical to the target assemblage $\{\{|\psi_{xk}\>\<\psi_{xk}|/d_\rmA\}_k\}_x$.
\end{proof}

\begin{proof}[Proof of \thref{thm:MESassemblage}]
	Let $\{\Pi_{xk}\}_k$ be the POVM of Bob used to generate the  ensemble $\{|\psi_{xk}\>\<\psi_{xk}|/d_\rmA\}_k$. Then we have  
	\begin{gather}
	\tr_\rmB[\rho (1_\rmA\otimes \Pi_{xk})]=\frac{1}{d_\rmA}|\psi_{xk}\>\<\psi_{xk}|,\\
	\sum_k \tr[\rho(|\psi_{x k}\>\<\psi_{x k}| \otimes \Pi_{x k} )]=1,\quad x=1,2,\ldots, m,
	\end{gather} 
	so $\rho$ is a ZUS with respect to $\scrB$, and $\{\Pi_{xk}\}_k$ is an optimal POVM for Bob. Thanks  to \thref{thm:ZUS},  we have $d_\rmB\geq d_\rmA$ and $\rho$ is a MES.  
\end{proof}
\begin{remark}	
	Given the state $\rho$, it is worth pointing out that the set of POVMs  required by Bob to generate the  assemblage $\{\{|\psi_{xk}\>\<\psi_{xk}|/d_\rmA\}_k\}_x$ is unique if we only consider the support of $\rho_\rmB$. This fact follows from  a similar argument presented  in Supplementary note E. 
\end{remark}

\subsection*{Supplementary note G: Proofs of \thref{thm:ZUSEB} and \crref{cor:ZUSEF1}}

\begin{proof}[Proof of \thref{thm:ZUSEB}]
	The first inequality in \eref{eq:ZUSEB} in \thref{thm:ZUSEB} follows from the fact that $\rho$ can be turned into $\rho_\scrB$ under LOCC.

	To prove the second inequality  in \eref{eq:ZUSEB} in \thref{thm:ZUSEB}, 
	note that each $\rho_a/q_a$ with $q_a>0$ is a MES on $\caH_{\rmA,a}\otimes \caH_\rmB$ and can be turned into a pure MES on $\caH_{\rmA,a}\otimes \caH_\rmB$ reversibly under local operations of Bob (cf. \thref{thm:ZUS2} and \crref{cor:MES} in the main text), so  
	\begin{equation}
	E(\rho_a/q_a)=E(|\Phi(r_a)\rangle).
	\end{equation}
	In addition, by the quantum operation $\Lambda_\scrB$ defined in \eref{eq:PhiB} in the Methods section, $\rho_\scrB$ can be turned into $\rho_a/q_a$ with probability $q_a$ (assuming $q_a>0$). Since the entanglement measure $E$ is nonincreasing on average under LOCC,  we have
	\begin{align}
	E(\rho_\scrB)&\geq \sum_{a| q_a>0}  q_a E(\rho_a/q_a)= \sum_{a| q_a>0} q_a E(|\Phi(r_a)\rangle)\nonumber\\
	&= \sum_a  q_a E(|\Phi(r_a)\rangle),
	\end{align}
	which confirms the second inequality in \eref{eq:ZUSEB} in \thref{thm:ZUSEB}.
	
	If 	$E$ is a convex	 entanglement measure $E$, then 
	\begin{equation}
	E(\rho_\scrB)\leq \sum_{a| q_a>0}  q_a E(\rho_a/q_a)=\sum_a  q_a E(|\Phi(r_a)\rangle),
	\end{equation}
	which implies that 
	\begin{align}
	E(\rho_\scrB)&=\sum_a  q_a E(|\Phi(r_a)\rangle)
	\end{align}
	in view of the above equation. So the second inequality in \eref{eq:ZUSEB} in \thref{thm:ZUSEB} is saturated in this case.
	
	The last inequality in \eref{eq:ZUSEB} in  \thref{thm:ZUSEB}
	follows from the normalization condition $\sum_a q_a=1$ and the simple fact that $E(|\Phi(r_a)\rangle)\geq E(|\Phi(r_{\min})\rangle) $, given that $|\Phi(r_a)\rangle$ can be turned into $|\Phi(r_{\min})\rangle$ under LOCC according to the majorization criterion \cite{Niel99}.
\end{proof}

\begin{proof}[Proof of \crref{cor:ZUSEF1}]
	The bound $E_\rmF(\rho)\geq\sum_a  q_a \log_2 r_a$ in \crref{cor:ZUSEF1} follows from \eref{eq:ZUSEB} in \thref{thm:ZUSEB} given that $E_\rmF(|\Phi(r_a)\rangle)=\log_2 r_a$. When $\rho=\rho_\scrB$,  the bound is saturated  since the first two inequalities in \eref{eq:ZUSEB} in \thref{thm:ZUSEB} are saturated. 
	
	To prove the converse,  let $\rho=\sum_s\alpha_s|\Psi_s\>\<\Psi_s|$ with $\alpha_s>0$ and $\sum_s \alpha_s=1$ be an optimal  convex  decomposition of $\rho$ that satisfies  $E_\rmF(\rho)=\sum_s \alpha_s E_\rmF(|\Psi_s\>)$. Then  each state $|\Psi_s\rangle$ is a ZUS according to \pref{pro:ZUSsupport} in the main text, and $\tr_\rmB(|\Psi_s\>\<\Psi_s|)$ is a weighted sum of component projectors in $\scrP(\scrB)$ according to \crref{cor:ZUSpure} in the main text. Therefore,
	\begin{align}
	E_\rmF(|\Psi_s\>)\geq E_\rmF(\Lambda_\scrB(|\Psi_s\>\<\Psi_s|))=\sum_a q_{s,a}\log_2 r_a,
	\end{align}
	where $\Lambda_\scrB$ is the CPTP map defined in \eref{eq:PhiB} in the Methods section and  $(q_{s,a})_a$ is the component vector of $|\Psi_s\rangle$ (cf. \thref{thm:ZUSEB}). Consequently, 
	\begin{align}
	E_\rmF(\rho)&=\sum_s \alpha_s E_\rmF(|\Psi_s\>)\geq \sum_s \alpha_s E_\rmF(\Lambda_\scrB(|\Psi_s\>\<\Psi_s|))\nonumber\\
	&=\sum_s \alpha_s \sum_a q_{s,a}\log_2 r_a=\sum_a q_a\log_2 r_a, \label{seq:EFineq}
	\end{align}
	where the last equality is due to the fact $q_a=\sum_s \alpha_s q_{s,a}$.
	
	If the inequality $E_\rmF(\rho)\geq \sum_a q_a \log_2 r_a$ is saturated, then the inequality in Supplementary \eref{seq:EFineq} is also saturated, which implies that 
	\begin{equation}
	E_\rmF(|\Psi_s\>)= E_\rmF(\Lambda_\scrB(|\Psi_s\>\<\Psi_s|)) \quad \forall s.
	\end{equation}
	So  $\Lambda_\scrB(|\Psi_s\>\<\Psi_s|)=|\Psi_s\>\<\Psi_s|$ and each reduced density operator $\tr_\rmB(|\Psi_s\>\<\Psi_s|)$ is supported on a component subspace, which implies that $\rho=\rho_\scrB$. This observation completes the proof of \crref{cor:ZUSEF1} in the main text.
\end{proof}
\newpage

\end{document}